\documentclass[12pt, a4paper]{article}
\usepackage{amsmath,amssymb}
\usepackage{graphicx}
\usepackage{color}
\usepackage{cite,fancyhdr}
\usepackage[affil-it]{authblk}

\begin{document}

\renewcommand{\thepage}{\arabic{page}}
\setcounter{page}{1}
\renewcommand{\thefootnote}{\#\arabic{footnote}}
\setcounter{footnote}{0}
\newcommand{\ubl}{U(1)$_{\rm B\,\mathchar`- L}$ }

\begin{titlepage}

\begin{center}

\hfill UT-18-25, TU-1075,  IPMU18-0177\\

\vskip .75in

{\Large \bf 
Revisiting the Number-Theory Dark Matter Scenario \\ \vspace{2mm} and the Weak Gravity Conjecture
}

\vskip .75in

{\large Kazunori Nakayama$^{a,b}$, Fuminobu Takahashi$^{c,b}$  and \\ Tsutomu T. Yanagida$^{b,d,e}$}

\vskip 0.25in

$^{a}${\em Department of Physics, Faculty of Science,\\
The University of Tokyo,  Bunkyo-ku, Tokyo 113-0033, Japan}\\[.3em]
$^{b}${\em Kavli Institute for the Physics and Mathematics of the Universe (WPI),\\
The University of Tokyo,  Kashiwa, Chiba 277-8583, Japan}\\[.3em]
$^{c}${\em Department of Physics, Tohoku University,
Sendai, Miyagi 980-8578, Japan}\\[.3em]
$^{d}${\em 
T.D.Lee Institute and School of Physics and Astronomy, \\ Shanghai Jiao Tong University, Shanghai 200240, China}\\[.3em]
$^{e}${\em Hamamatsu Professor}

\end{center}
\vskip .5in

\begin{abstract}
We revisit the number-theory dark matter scenario where one of the light chiral fermions required by the anomaly cancellation conditions of \ubl explains dark matter. 
Focusing on some of the integer ${\rm B\,\mathchar`- L}$ charge assignments,
we explore a new region of the parameter space where there appear two light fermions
and the heavier one becomes a dark matter of mass $\lesssim {\cal O}(10)$\,keV or $ {\cal O}(10)$\,MeV.
The dark matter radiatively decays into neutrino and photon, which can explain the tantalizing 
hint of the  $3.55$\,keV X-ray line excess.
Interestingly, the other light fermion can erase the AdS vacuum around the neutrino mass scale in a compactification of the standard model to 3D. 
This will make the standard model consistent with the AdS-WGC statement that stable non-supersymmetric AdS vacua should be absent. 
\end{abstract}

\end{titlepage}



\newpage

\section{Introduction}
\label{sec:1}

The seesaw mechanism is the most attractive mechanism to explain small masses of the active neutrinos~\cite{Yanagida:1979as,GellMann:1980vs,Minkowski:1977sc}. It is based on a new gauge symmetry called  U(1)$_{\rm B\,\mathchar`- L}$, which requires an extension of the standard model (SM) to make \ubl anomaly-free. The \ubl symmetry becomes anomaly-free if one introduces a right-handed neutrino (RHN) in each generation.
The spontaneous breaking of the \ubl  generates large Majorana masses for the RHNs, and integrating them out induces small Majorana masses for the active neutrinos through the seesaw mechanism.  Not only does it explain the smallness of neutrino masses, but it can also explain the present baryon asymmetry via leptogenesis in which
      lepton asymmetry is generated by the decay of the RHNs in the early Universe~\cite{Fukugita:1986hr}. 

In a conventional framework of the SM plus three RHNs,  the  RHNs acquire a heavy 
mass of order the \ubl breaking scale, and there is no dark matter (DM) candidate.
 In the so-called split seesaw mechanism~\cite{Kusenko:2010ik}, on the other hand, one of the RHNs becomes much lighter than the others. If the lightest one has a mass less than
     $\mathcal O(10){\rm\, keV}$ and if its Yukawa couplings are sufficiently suppressed to satisfy the X-ray bound, 
     it becomes a plausible candidate for DM. Note that, while the lightest RHN is almost decoupled from the others in this case,  the seesaw mechanism, as well as leptogenesis, still work thanks to the remaining two heavy RHNs~\cite{Frampton:2002qc,Endoh:2002wm}. The split seesaw mechanism provides a possible answer to the question of why there are three generations in the SM: one of the three RHNs becomes DM, while the other two explain the baryon asymmetry of the Universe via leptogenesis.
     
     Alternatively, one could introduce a set of chiral fermions charged under \ubl in addition to the three RHNs. The number of such chiral fermions and their B-L charges are subject to the anomaly cancellation conditions, and
      it turned out that their number must be greater than or equal to five, partly because of the absence of positive integers satisfying $x^3 + y^ 3 = z^3$, a special case of the Fermat's last theorem~\cite{Wiles:1995ig}. 
      Interestingly,  the lightest extra chiral fermion with even B-L charge is stable and therefore a candidate for DM. Since there is an intimate connection between the number theory and the existence of DM through the anomaly cancellation condition, we named the above model
       as the number-theory dark matter (NTDM)~\cite{Nakayama:2011dj}.\footnote{
    See also Refs.~\cite{Montero:2007cd,Sanchez-Vega:2014rka,Ma:2014qra,Sanchez-Vega:2015qva,Patra:2016ofq,Bernal:2018aon} for related works on the possible extensions of the SM with 
    \ubl charged chiral fermions.}  

In this Letter, we revisit the NTDM model and explore a new region of the parameter space where there appear two light chiral fermions in the low energy. Specifically,  we will focus on two possible sets of extra fermions with integer ${\rm B\,\mathchar`- L}$ charges; one is equivalent to the case with four RHNs and four extra chiral fermions, while the other is to the case with two RHNs and four extra chiral fermions. 
The latter case can be thought of as a realization of the split seesaw mechanism, where one RHN is traded with four chiral fermions that give the equivalent contributions to the anomalies. 
In both cases, we will show that there is a quasi-stable fermion in addition to an ultralight stable fermion. 

The quasi-stable fermion can be sufficiently long-lived and explain DM 
if its mass is sufficiently light.
While it mainly decays into three neutrinos, it also radiatively decays into neutrino and photon.
Such decaying DM can be searched for in astrophysical/cosmological X-ray observations. 
For a certain choice of the parameters,
it can explain the 3.55 keV X-ray line excess~\cite{Bulbul:2014sua,Boyarsky:2014ska,Boyarsky:2014jta}. 

Intriguingly,  our set-up contains not only a DM candidate but also one ultralight chiral fermion, whose existence may help the SM to be consistent with quantum gravity. 
  Recently, it was argued based on the sharpened version of the weak gravity conjecture (WGC)~\cite{ArkaniHamed:2006dz,Ooguri:2016pdq} that non-supersymmetric (SUSY) anti-de-Sitter (AdS) potential minimum should not exist in order for the low-energy effectively theory to be successfully UV completed into a theory of quantum gravity. As shown in Ref.~\cite{ArkaniHamed:2006dz}, the SM compactified on a circle has such an AdS minimum around the neutrino mass scale in the one-loop effective potential for the radion field, if the light neutrino mass is of Majorana type as predicted in the seesaw mechanism. 
  The presence of the AdS minimum around the neutrino mass scale in the compactified SM crucially depends on the number of extra particles as light as active neutrinos. As we shall see, 
  the ultralight fermion in our NTDM model can erase the unwanted AdS minimum. 

The rest of this Letter is organized as follows. In Sec.~\ref{sec:NTDM}, after briefly reviewing the NTDM model, 
we study the properties of the light fermions for two possible sets of extra fermions with integer ${\rm B\,\mathchar`- L}$ charges. 
In Sec.~\ref{sec:3} we discuss cosmological aspects of our scenario such as the production mechanism of DM and leptogenesis. In Sec.~\ref{sec:AdS} we will show that the ultralight fermion helps to erase the AdS minimum at the neutrino mass scale. The last section is devoted to  conclusions. 

\section{Number-Theory Dark Matter} \label{sec:NTDM}

The central ingredient of the NTDM model is the two anomaly cancellation conditions of 
U(1)$_{\rm B\,\mathchar`- L}$; one is $[{\rm U(1)}_{\rm B\,\mathchar`- L}]^3$ anomaly and
the other is the gravitational $[{\rm U(1)}_{\rm B\,\mathchar`- L}]\times[graviton]^2$
anomaly. Both anomalies are absent in the SM plus three RHNs, but there is no DM
candidate if all the RHNs are heavy. 
This suggests that some extension is needed.

Let us introduce $n$ left-handed Weyl fermions $\psi_i$ with the ${\rm B\,\mathchar`- L}$
charge $q_i$. Then the two anomaly cancellation conditions read
\begin{align}
    \sum_{i=1}^{n} q_i &= 0,~~~~~~\sum_{i=1}^{n} q_i^3 = 0.  \label{anomalyfree}
\end{align}
Here we restrict ourselves to integer (or rational) number solutions to the above equations~\cite{Banks:2010zn}. 
We also exclude a vector-like charge assignment since a vector-like
pair such as $\psi(+1)$ and $\psi(-1)$ would acquire a large mass close to the cut-off scale of the theory and therefore become irrelevant for the low-energy physics. 

In Table~\ref{table:ntdm} we show the first several integer solutions with $n=5$~\cite{Batra:2005rh,Nakayama:2011dj}.
In fact, there are no solutions with $n = 2,3,$ and 4. The cases of $n=2$ and 4 lead to
only vector-like charge assignments. In the case of $n=3$, there are no positive integers satisfying $x^3 + y^3 = z^3$, which is nothing but a special case of the Fermat's last theorem.

\begin{table}
\caption{Examples of the integer solutions to the conditions (\ref{anomalyfree})
in the case of $n=5$.} 
\begin{center}
\begin{tabular}{ c c c c c} \hline
    $q_1$ & $q_2$ & $q_3$ & $q_4$ & $q_5$ \\ \hline\hline
    $-9$ & $-5$ & $-1$ & $7$ & $8$  \\ \hline
    $-9$ & $-7$ & $2$ & $4$ & $10$  \\ \hline
    $-18$ & $-17$ & $1$ & $14$ & $20$  \\ \hline
    $-21$ & $-12$ & $5$ & $6$ & $22$  \\ \hline
    $-25$ & $-8$ & $-7$ & $18$ & $22$  \\ \hline
\end{tabular}
\label{table:ntdm}
\end{center}
\end{table}

There are several remarks on these solutions. First, as emphasized in~\cite{Nakayama:2011dj}, 
a charge assignment obtained by multiplying the integer solutions in Table~\ref{table:ntdm} with an overall rational number also satisfies the anomaly cancellation conditions. 
By doing so, we can obtain fractional charge solutions.\footnote{
The fractional charge solutions generically contain massless fermions, which can erase the local AdS minimum around the neutrino 
mass scale, similarly to the discussion
in Sec.~\ref{sec:AdS}.
} Secondly, by using this freedom of the overall normalization of the ${\rm B\,\mathchar`- L}$ charge, we can always make one of the Weyl fermions to have a ${\rm B\,\mathchar`- L}$ charge of $-1$. 
    The fermion $\psi(-1)$ forms a vector-like mass with a linear combination of the RHNs, and therefore they can be integrated out.  In such a case the other four fermions satisfy
\begin{align}
    \sum_{i=1}^{4} q_i &= 1,~~~~~~\sum_{i=1}^{4} q_i^3 = 1.
\end{align}
In other words, it is possible to trade one of the RHNs with the four chiral fermions satisfying the
above conditions, and then we are left with two RHNs plus additional four Weyl fermions. 

The ${\rm B\,\mathchar`- L}$ Higgs $\Phi$ develops a nonzero vacuum expectation value (VEV),
 $v_\Phi = \langle \Phi \rangle$, which is taken to be real and positive in the unitary gauge. 
 Note that there is a residual $Z_2^{\rm B\,\mathchar`- L}$ symmetry, which restricts the mixings among
 the fermions depending on their ${\rm B\,\mathchar`- L}$ charges. 
 Hereafter we consider two possibilities based on the first solution of Table~\ref{table:ntdm}. 
The ${\rm B\,\mathchar`- L}$ charge assignments of the relevant fields are summarized in Table~\ref{table}, 
where $L_a$ and $N_I$ represent the lepton doublets and the RHNs, respectively.

\begin{table}
\caption{The ${\rm B\,\mathchar`- L}$ charge assignments in the models A and B
} 
\begin{center}
\begin{tabular}{c c c c c c c c c c c c c} 
\hline
    ~ &$L_1$ & $L_2$ & $L_3$ & $N^c_1$ & $N^c_2$ & $N^c_3$ & $\psi_1$ & $\psi_2$  & $\psi_3$ & $\psi_4$ & $\psi_5$ & $\Phi$ \\ \hline\hline
    Model A & $-1$ & $-1$ & $-1$ & $1$ & $1$ & $1$ & $9$  & $5$ & $1$ & $-7$ & $-8$ & $-2$ \\
    Model B & $-1$ & $-1$ & $-1$ & $1$ & $1$ & - & $-9$  & $-5$ & - & $7$ & $8$ & $-2$ \\
    \hline
\end{tabular}
\label{table}
\end{center}
\end{table}

\subsection{Model A}

Let us consider the model A which contains three RHNs as well as five Weyl fermions $\psi_1,\dots, \psi_5$~\cite{Nakayama:2011dj}. 
See Table~\ref{table} for the charge assignment. 

The active neutrino masses are generated through the seesaw mechanism. 
Noting that $\psi_3$ has the same ${\rm B\,\mathchar`- L}$ charge as the RHNs, 
we define $N_4^c\equiv \psi_3$ for the notational convenience. Then we have
\begin{align}
    \mathcal L = \frac{1}{2} y_N^{IJ} \Phi N_I^c N_J^c + y_\nu^{aI} L_a N^c_I H + {\rm h.c.}  \label{seesaw}
\end{align}
where  $H$ is the SM Higgs doublet. The RHNs obtain masses of order $v_\Phi$ after $\Phi$ develops a VEV,
and integrating them out leads to the small neutrino masses. Although the active neutrinos are also 
mixed with other fermions, their contributions to the neutrino mass matrix are highly suppressed 
as discussed below. 

The other fermions become massive after the ${\rm B\,\mathchar`- L}$ breaking.
Let us first consider $\psi_5$,  which is the only fermion 
that has an even ${\rm B\,\mathchar`- L}$ charge. Due to the residual $Z_2^{\rm B\,\mathchar`- L}$ symmetry,
it does not mix with the other fermions, and so, its mass arises only from the 
high dimensional operator,
\begin{align}
    \mathcal L = \frac{\Phi^{8}}{2M^7} \psi_5 \psi_5 + {\rm h.c.},
\end{align}
where $M$ denotes the cutoff scale. The mass of $\psi_5$ is given by
\begin{align}
    m_{\psi_5} = \frac{v_\Phi^8}{M^7} \simeq 7.7\times 10^{-11}\,{\rm eV} \left( \frac{v_\Phi}{5\times 10^{13}\,{\rm GeV}} \right)^8\left( \frac{M_P}{M} \right)^7,
    \label{m5}
\end{align}
where
 $M_P\simeq 2.4\times 10^{18}\,{\rm GeV}$ is the reduced Planck mass. 
 Note that $\psi_5$ is stable due to the $Z_{2}^{\rm B\,\mathchar`- L}$ symmetry 
regardless of the mass.

Next, let us consider other fermions $\psi_i$ $(i=1,\dots,4)$ and RHNs, which have mixings with one another.
As mentioned above, the RHNs obtain heavy masses of order $v_\Phi$.
The other fermions, $\psi_1$, $\psi_2$ and $\psi_4$, are split into heavy and light modes. We can write down the mass term as
\begin{align}
    \mathcal L = y_1\Phi\psi_1 \psi_4 +  y_2\Phi^*\psi_2 \psi_4 + {\rm h.c.}, 
\end{align}
where $y_1$ and $y_2$ are numerical coefficients of order unity, and we take them real 
and positive without loss of generality. 
Thus one combination of $\psi_1$ and $\psi_2$ denoted by $\psi_h$ 
forms a heavy Dirac mass with $\psi_4$, while
the other combination $\psi_\ell$ remains light. They are  given by
\begin{align}
    \psi_h \equiv \frac{y_1 \psi_1+ y_2 \psi_2}{\sqrt{y_1^2+y_2^2}},~~~~~~\psi_\ell \equiv \frac{-y_2 \psi_1+ y_1 \psi_2}{\sqrt{y_1^2+y_2^2}}.
\end{align}
The $\psi_h$ and $\psi_4$ have a heavy Dirac mass of $\sim v_\Phi$. 
To estimate the mass of $\psi_\ell$, let us explicitly write down the mass matrix of $\psi_1$, $\psi_2$, $\psi_4$ and $N_I$ $(I=1,\dots,4)$:
\begin{align}
    \mathcal L = \frac{1}{2} (\psi_1,\psi_2,\psi_4,N_I) \mathcal M \begin{pmatrix}
        \psi_1 \\ \psi_2 \\ \psi_4 \\ N_I
    \end{pmatrix} + {\rm h.c.},~~~
    \mathcal M \simeq v_\Phi \begin{pmatrix}
        \epsilon^8 & \epsilon^6 &  \epsilon^0 & \epsilon^4 \\
        \epsilon^6 & \epsilon^4 &   \epsilon^0  & \epsilon^2\\
          \epsilon^0 &   \epsilon^0  & \epsilon^6 & \epsilon^4 \\
        \epsilon^4 & \epsilon^2 & \epsilon^4 &   \epsilon^0 
    \end{pmatrix}.
\end{align}
where we have defined $\epsilon \equiv (v_\Phi / M)$ and omitted numerical coefficients of order unity. 
It is found that there is one light mode $\psi_\ell$ whose mass eigenvalue is
\begin{align}
    m_{\psi_\ell} \simeq \frac{v_\Phi^5}{M^4} \simeq  8.9\,{\rm keV} \left( \frac{v_\Phi}{5\times10^{13}\,{\rm GeV}} \right)^5\left( \frac{M_P}{M} \right)^4. 
\end{align}
To summarize, there are two light chiral fermions, $\psi_5$ and $\psi_\ell$, in our model. The other fermions are as heavy as the ${\rm B\,\mathchar`- L}$ breaking scale $\sim v_\Phi$. Thus we will consider phenomenological aspects of these light fermions in the following.

In Ref.~\cite{Nakayama:2011dj} the lightest stable fermion $\psi_5$ was identified with DM. 
Here we explore a new region of the parameter space where $\psi_\ell$  
plays a role of DM. 
Even though the stability of $\psi_\ell$ is not guaranteed by symmetry in contrast to $\psi_5$, it can be sufficiently
long-lived as we shall see below.

The $\psi_\ell$ has a mixing with active neutrinos through the operator
\begin{align}
    \mathcal L \simeq \kappa_{a}\left( \frac{\Phi^{*}}{M} \right)^2 \psi_2 L_a H + {\rm h.c.}.   \label{psi2LH}
\end{align}
The typical value of the mixing angle between $\psi_\ell$ and active neutrino is given by
\begin{align}
    \theta \simeq \frac{\kappa \epsilon^2 v_H}{m_{\psi_\ell}} \sim 8\times 10^{-3}\,\kappa \left( \frac{5\times10^{13}\,{\rm GeV}}{v_\Phi} \right)^3\left( \frac{M}{M_P} \right)^2, \label{thetaA}
\end{align}
where $\kappa$ represents the largest value among $\kappa_{a}$.
The decay modes depend on the mass of $\psi_\ell$. If it is lighter than $\sim 1\,$MeV, it dominantly decays into active neutrinos as $\psi_\ell \to \nu\nu \bar\nu$ with a decay width of~\cite{Pal:1981rm,Adhikari:2016bei}
\begin{align}
    \Gamma_{\psi_\ell \to 3\nu} &= \theta^2 \frac{G_F^2 m_{\psi_\ell}^5}{96\pi^3}
    \simeq (3.8\times 10^{18}\,{\rm sec})^{-1}\,\kappa^2
    \left( \frac{v_\Phi}{5\times10^{13}\,{\rm GeV}}\right)^{19}\left( \frac{M_P}{M} \right)^{16}.
\end{align}
where $G_F$ denotes the Fermi constant. For $1\,{\rm MeV} \lesssim m_{\psi_\ell} \lesssim 210$\,MeV, it also decays into  $\nu e^+e^-$ and the total decay width becomes~\cite{Ruchayskiy:2011aa}
\begin{align}
    \Gamma_{\psi_\ell} \simeq \left(1+ g_{L}^2 + g_R^2\right)\theta^2\frac{G_F^2 m_{\psi_\ell}^5}{96\pi^3},
\end{align}
where $g_R=\sin^2\theta_W$ and $g_L=1/2\pm \sin^2\theta_W$, with $\theta_W$ being the weak mixing angle.
The plus (minus) sign corresponds to the case where $\psi_\ell$ is dominantly mixed with the electron-type (mu- or tau-type) active neutrino.
For even heavier $\psi_\ell$, new decay modes into heavier leptons and hadrons will open, but here we do not consider this case further since the lifetime tends to become much shorter than the present age of the Universe and it is not suitable for DM.
On the other hand, due to the Tremaine-Gunn bound~\cite{Tremaine:1979we}, any fermionic DM particle cannot be much lighter than $\sim 1$\,keV. 
Thus the interesting mass range is $1\,{\rm keV} \lesssim m_{\psi_\ell} \lesssim 1$\,MeV.

\begin{figure}
    \begin{center}
        \includegraphics[scale=1.3]{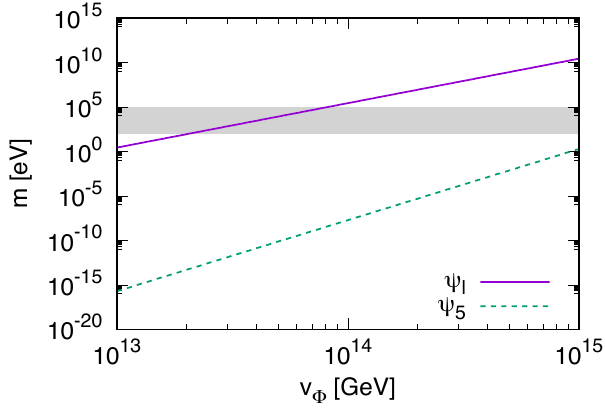}
        \includegraphics[scale=1.3]{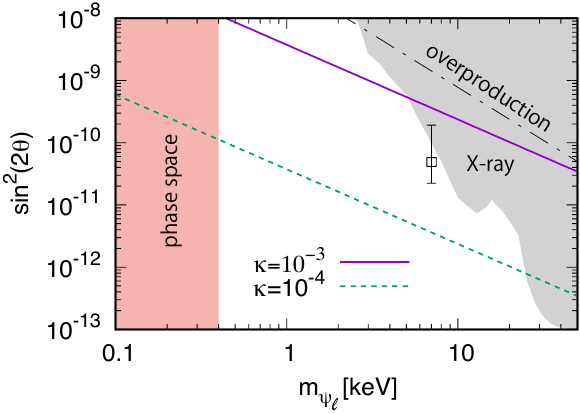}
        \caption{Left: Mass of $\psi_\ell$ (upper) and $\psi_5$ (lower) as a function of $v_\Phi$. 
        We are interested in the mass range of $\psi_\ell$ indicated by the gray band.
        Right: Mixing angle $\sin^2(2\theta)$ as a function of $m_{\psi_\ell}$ for $\kappa=10^{-3}$ (upper) and $10^{-4}$ (lower).
        The gray shaded region and the region above the dot-dashed line are excluded by X-ray observations
        and the DM overabundance, respectively. We have taken $M=M_P$ in both plots.
        The black data point with an error bar shows the parameter region where 3.55\,keV excess may be explained by decaying DM. 
        \label{fig:A}}
    \end{center}
\end{figure}

For such a mass range of $\psi_\ell$, it can also decay into active neutrino plus photon through a one-loop diagram. The partial decay rate of this radiative process is given by~\cite{Pal:1981rm}
\begin{align}
    \Gamma_{\psi_\ell \to \nu\gamma} = \theta^2 \frac{9 \alpha_e G_F^2 m_{\psi_\ell}^5}{256\pi^4},  \label{nugamma}
\end{align}
with $\alpha_e$ being the electromagnetic fine-structure constant. For a keV-scale DM, 
this may make a significant contribution to the measured X-ray fluxes. For DM mass 
of $1$--$10$\,keV, the observational constraint is roughly 
$\theta^2 \lesssim 10^{-8}$--$10^{-11}$~\cite{Adhikari:2016bei}. 
There is a tantalizing hint for the X-ray line excess at $3.55$\,keV, which may be interpreted as 
the decay of DM of mass about $7.1$\,keV such as sterile neutrino~\cite{Ishida:2014dlp} 
or axion~\cite{Higaki:2014zua,Jaeckel:2014qea}. Specifically, the excess can be explained 
by sterile neutrino DM of mass $\simeq 7.1$\,keV and the mixing $\theta \simeq (2-7)\times 10^{-6}$.

Fig.~\ref{fig:A} shows the mass of $\psi_\ell$ and $\psi_5$ as a function of $v_\Phi$, and $\sin^2(2\theta)$ as a function of $m_{\psi_\ell}$ for $\kappa=10^{-3}$ (upper) and $10^{-4}$ (lower). 
In the left panel, the horizontal gray band shows $m_{\psi_\ell}= 1 - 100$\,keV, 
where $\psi_\ell$ is a good candidate for DM. In the right panel,
the gray shaded region is excluded by X-ray observations~\cite{Adhikari:2016bei},
and the region above the dot-dashed line is excluded by the overproduction of $\psi_\ell$ through the Dodelson-Widrow
mechanism~\cite{Dodelson:1993je}.
It is seen that  the constraint from X-ray observation is satisfied for $m_{\psi_\ell} = {\cal O}(1-10)$\,keV and $\kappa 
\lesssim {\cal O}(10^{-3})$. 
Such relatively small values of the Yukawa couplings $\kappa_{a}$ in (\ref{psi2LH}) may be explained if $L_a$ and/or $\psi_2$ are charged under a global Abelian flavor symmetry and Yukawa couplings arise as a result of its spontaneous breaking as in the Froggatt-Nielsen mechanism~\cite{Froggatt:1978nt,Ishida:2013mva,Ishida:2014dlp}.

\subsection{Model B}
In the model B, we flip the sign of the extra Weyl fermions. Then, one of them forms a heavy Dirac mass with
a linear combination of the three RHNs which we take to be $N_3$ for simplicity. 
As a result,  the third RHN $N_3$ and one of the additional Weyl 
fermions $\psi_3$ are absent, compared to the model A.  Therefore, the particle content is more minimal than
the model A.

The active neutrino masses and mixings are dominantly generated by the seesaw 
mechanism with the two heavy RHNs $N_1$ and $N_2$:
\begin{align}
    \mathcal L = \frac{1}{2} y_N^{IJ} \Phi N_I^c N_J^c + y_\nu^{aI} L_a N^c_I H + {\rm h.c.}  \label{seesawB}
\end{align}
This is the same as Eq.~(\ref{seesaw})  in the model A, but now we  have only two RHNs $(I=1,2)$.
In the usual scenario with the two RHNs, the lightest active neutrino is massless~\cite{Frampton:2002qc}. 
In the present model,
however, it acquires a tiny mass through the mixings with the
other fermions $\psi_1,\psi_2$ and $\psi_4$. As we shall see below, the mixings are highly suppressed by powers of $(v_\Phi/M)$ because of the non-trivial \ubl charge assignments, and so, the lightest neutrino mass
is safely neglected for phenomenological purposes.

The mass of the fermion, $\psi_5$, is the same as given in Eq.~(\ref{m5}), and its stability is guaranteed by the $Z_2^{\rm B\,\mathchar`- L}$ and Lorentz symmetries. 
On the other hand,  $\psi_1$, $\psi_2$, $\psi_4$ and $N_I$ $(I=1,2)$ get mixed with one another. 
Similarly to the model A, one linear combination of $\psi_1$ and $\psi_2$, $\psi_h$, 
forms a heavy Dirac mass with $\psi_4$, while the other combination, $\psi_\ell$,
remains light. As a result,  the two RHNs, $\psi_h$, and $\psi_4$ acquire a mass of order $ v_\Phi$.
To estimate the mass of $\psi_\ell$, let us write down the mass matrix of $\psi_1$, $\psi_2$, $\psi_4$
 and $N_I$ as
\begin{align}
    \mathcal L = \frac{1}{2} (\psi_1,\psi_2,\psi_4,N_I) \mathcal M \begin{pmatrix}
        \psi_1 \\ \psi_2 \\ \psi_4 \\ N_I
    \end{pmatrix} + {\rm h.c.},~~~
    \mathcal M \simeq v_\Phi \begin{pmatrix}
        \epsilon^8 & \epsilon^6 & \epsilon^0 & \epsilon^3 \\
        \epsilon^6 & \epsilon^4 & \epsilon^0 & \epsilon\\
        \epsilon^0 & \epsilon^0 & \epsilon^6 & \epsilon^3 \\
        \epsilon^3 & \epsilon & \epsilon^3 & \epsilon^0
    \end{pmatrix}.
\end{align}
where we have omitted numerical coefficients of order unity. 
It is found that there is one light mode $\psi_\ell$ whose mass eigenvalue is
\begin{align}
    m_{\psi_\ell} \simeq \frac{v_\Phi^3}{M^2} \simeq  1.7\times 10^2\,{\rm keV} \left( \frac{v_\Phi}{10^{11}\,{\rm GeV}} \right)^3\left( \frac{M_P}{M} \right)^2. 
\end{align}
Therefore, again we are left with two light chiral fermions, $\psi_5$ and $\psi_\ell$, in the model B. 
The other fermions are as heavy as the ${\rm B\,\mathchar`- L}$ breaking scale $\sim v_\Phi$. 
In this respect, our model can be regarded as a natural realization of the split 
seesaw mechanism~\cite{Kusenko:2010ik} if we identify $\psi_\ell$ with the light sterile neutrino.
Note that we need a relatively low ${\rm B\,\mathchar`- L}$ breaking scale, $v_\Phi \sim 10^{11}$\,GeV, in order for
$\psi_\ell$ to be a DM candidate. As a result, the neutrino Yukawa couplings  in (\ref{seesawB}) should also be mildly suppressed as  $y_{\nu}^{aI} = \mathcal O(10^{-2})$ for the successful seesaw mechanism,
which may be explained by some broken U(1) symmetry.

\begin{figure}[t!]
    \begin{center}
        \includegraphics[scale=1.3]{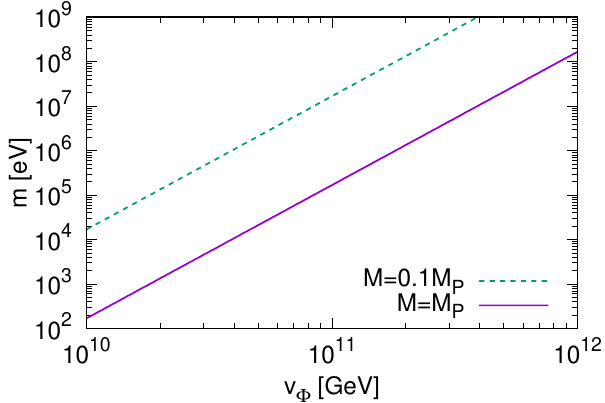}
        \includegraphics[scale=1.3]{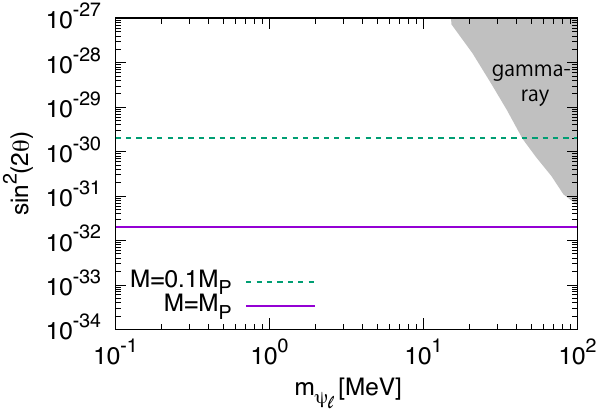}
        \caption{Left: Mass of $\psi_\ell$ as a function of $v_\Phi$ for $M=0.1M_P$ (upper) and $M_P$ (lower).
        Right: Mixing angle $\sin^2(2\theta)$ as a function of $m_{\psi_\ell}$ for $M=0.1M_P$ (upper) and $M_P$ (lower).
        The gray shaded region is excluded by X- and gamma-ray observations~\cite{Essig:2013goa}.  
        We have taken $\kappa=1$.
        \label{fig:B}}
    \end{center}
\end{figure}

The light fermion $\psi_\ell$ mixes with the active neutrinos through the operator
\begin{align}
    \mathcal L \simeq \kappa_{a}\left( \frac{\Phi^{*}}{M} \right)^3 \psi_2 L_a H + {\rm h.c.},
\end{align}
and it decays into the SM particles in a way that depends on the mass of $\psi_\ell$. 
As explained in the previous subsection, if $\psi_\ell$ is lighter than $\sim 1\,$MeV, 
it dominantly decays into active neutrinos as $\psi_\ell \to \nu\nu\bar\nu$ with a decay width of
\begin{align}
    \Gamma_{\psi_\ell \to 3\nu} &= \theta^2 \frac{G_F^2 m_{\psi_\ell}^5}{96\pi^3}
    \simeq (2.1\times 10^{40}\,{\rm sec})^{-1}\,\kappa^2
    \left( \frac{v_\Phi}{10^{11}\,{\rm GeV}}\right)^{15}\left( \frac{M_P}{M} \right)^{12}.
\end{align}
Here the mixing angle between $\psi_\ell$ and active neutrino is given by
\begin{align}
    \theta \simeq \frac{\kappa \epsilon^3 v_H}{m_{\psi_\ell}} \sim 7\times 10^{-17}\,\kappa \left( \frac{M_P}{M} \right),\label{thetaB}
\end{align}
which is independent of $v_\Phi$. 

The left panel of Fig.~\ref{fig:B} shows the mass of $\psi_\ell$ as a function of $v_\Phi$ for $M=0.1M_P$ (upper) and $M_P$ (lower).
The right panel of Fig.~\ref{fig:B} shows of $\sin^2(2\theta)$ as a function of $m_{\psi_\ell}$ for $M=0.1M_P$ (upper) and $M_P$ (lower) with $\kappa=1$.
 Note that $\psi_{\ell}$ also decays into $\nu$ plus $\gamma$ with the decay width given in Eq.~(\ref{nugamma}),
 and also into $\nu e \bar{e}$ if kinematically allowed. In particular, 
for $m_{\psi_\ell} = {\cal O}(100)$\,MeV, the predicted photon flux is close to the observational upper bound from the diffuse X- and gamma-ray flux by COMPTEL and EGRET~\cite{Essig:2013goa}. On the other hand, $\psi_5$ is extremely light, and it plays no cosmological role.
As we shall see in Sec.~\ref{sec:AdS}, however, the existence of such an ultralight fermion can ensure the absence of AdS minimum in the compactified SM.



\section{Dark Matter Production}
\label{sec:3}

So far we have seen that $\psi_\ell$ can have a suitable mass and lifetime so that 
it can be a candidate for DM. Now let us estimate the abundance of $\psi_\ell$. 
The $\psi_\ell$ has suppressed mixings with the active neutrinos, and so, it can be thought of
as a sterile neutrino. The production of sterile neutrino DM has been extensively studied
in the literature (see e.g. Refs.~\cite{Adhikari:2016bei,Boyarsky:2018tvu} for recent reviews 
and references therein). The sterile neutrino is necessarily produced through the so-called Dodelson-Widrow
(or non-resonant production) mechanism~\cite{Dodelson:1993je}. It is known, however, that one cannot
produce the right amount of DM through the mechanism while satisfying the X-ray constraints, because both
the production and decay rates are controlled by the mixing angle. 

The relevant production process in our set-up is the pair production of $\psi_\ell$ 
particles through the s-channel ${\rm B\,\mathchar`- L}$ gauge boson exchange from the scattering of SM particles 
in thermal bath~\cite{Khalil:2008kp,Nakayama:2011dj}. The resultant abundance, in terms of the number density 
$(n_{\psi_\ell})$ to the entropy density $(s)$ ratio, may be estimated as
\begin{align}
    \left[\frac{n_{\psi_\ell}}{s}\right]_{\rm gauge} \simeq 3\times 10^{-4}\,q_{\psi_\ell}^2 
    \left(\frac{10^2}{g_*}\right)^\frac{3}{2}
    \left(  \frac{5\times10^{13}\,{\rm GeV}}{v_\Phi} \right)^4\left( \frac{T_{\rm R}}{10^{12}\,{\rm GeV}} \right)^3,  \label{Y_B-L}
\end{align}
where $g_*$ denotes the effective relativistic species at the reheating, 
$T_{\rm R}$ denotes the reheating temperature after inflation and
\begin{align}
    q_{\psi_\ell} = \frac{-y_2 q_1 + y_1 q_2 }{\sqrt{y_1^2+y_2^2}}
\end{align}
denotes the effective charge of $\psi_\ell$. In terms of the density parameter, the $\psi_\ell$ abundance 
is given by
\begin{align}
\Omega_{\psi_\ell}h^2 \sim 0.1\,
q_{\psi_\ell}^2 
 \left(\frac{m_{\psi_\ell}}{1 {\rm\,keV}}\right)
    \left(\frac{10^2}{g_*}\right)^\frac{3}{2}
    \left(  \frac{5\times10^{13}\,{\rm GeV}}{v_\Phi} \right)^4\left( \frac{T_{\rm R}}{10^{12}\,{\rm GeV}} \right)^3.
\end{align}
%
%
%
The observed DM abundance can be explained for $T_{\rm R} \sim 10^{12}\,$GeV in the model A,
 and for $T_{\rm R} \sim 10^8$\,GeV in the model B, respectively. In this case, the thermal leptogenesis works in model A,  but one needs to assume non-thermal and/or resonant leptogenesis in model B.

Note that $\psi_5$ is also produced in a similar way, and its abundance can be estimated by 
replacing $q_{\psi_\ell}$ with $q_5$.
As is clear from (\ref{Y_B-L}), $\psi_5$ is not likely to be thermalized. 
Thus its contribution to the effective number of neutrino species is much smaller than unity, 
so that it only has negligible effects on cosmological observations.

\section{Erasing the AdS vacuum at the neutrino mass scale} 
 \label{sec:AdS}

Our NTDM scenario may also have an interesting implication for the conjecture proposed in Ref.~\cite{Ooguri:2016pdq},
which states that the existence any non-SUSY AdS vacuum is not consistent with quantum gravity.
If this is true, the SM with three species of light Majorana neutrino is ruled out, since there appears a stable 
AdS vacuum in the SM compactified on a circle~\cite{ArkaniHamed:2007gg,Ibanez:2017kvh}.

Let us suppose that one of the spatial directions (say, $z$ direction) is compactified on a circle with radius $R$. The radius of the compactified dimension, which we call radion, is a dynamical field and it obtains a potential at the one-loop level as a Casimir energy arising from the SM field as well as the tree-level effect from the 4D cosmological constant. The Casimir energy density for the massive field with species $i$ is given by
\begin{align}
    \rho_i(R) = \mp g_i \left[- \int \frac{d^2 k_{\parallel}}{(2\pi)^2}\left( \frac{\omega_0(k_\parallel)}{2} + \sum_{n\geq 1}\omega_n(k_\parallel) \right)
    +(2\pi R) \int \frac{d^3 k}{(2\pi)^3} \frac{\omega_k}{2} \right],
\end{align}
where the minus (plus) sign corresponds to the boson (fermion), $g_i$ denotes the degree of freedom ($1$ for a real scalar, $2$ for a Weyl fermion, and so on), $m_i$ denotes its mass,\footnote{
    For simplicity, we imposed periodic boundary conditions for all fields in the $z$ direction.
} and
\begin{align}
    \omega_n^2(k_\parallel) = k_\parallel^2 + \left(\frac{n}{R}\right)^2+m_i^2,~~~~~~\omega_k^2 = k^2_{\parallel} + k_z^2 + m_i^2.
\end{align}
It is analytically calculated in several ways as~\cite{Ambjorn:1981xw,Plunien:1986ca}
\begin{align}
    \rho_i(R) =\mp g_i \sum_{n=1}^{\infty} \frac{2m_i^4}{(2\pi)^4} \frac{K_2(2\pi R m_i n)}{(2\pi R m_i n)^2},
\end{align}
where $K_2(x)$ is the modified Bessel function of the second kind with order $2$.

For a massless field $(m_i=0)$, it becomes
\begin{align}
    \rho_i(R) = \mp \frac{g_i}{1440 \pi^2 R^4}.
\end{align}
It is checked that for a massive field, the contribution to the Casimir energy is exponentially suppressed as $e^{-2\pi m_i R}$.
Since we are interested in the potential structure around the neutrino mass scale, we will consider only the light fields, i.e., photon, graviton, neutrinos and the ultralight fermion in our NTDM model.

After including the effect of cosmological constant on the Casimir energy, the effective 3D radion potential is  given by
\begin{align}
    V(R) = \frac{2\pi r^3 \Lambda_4}{R^2} + \sum_i \frac{2\pi r^3}{R^2}\rho_i(R),
\end{align}
where $\Lambda_4$ is the observed 4D cosmological constant. Here we have used the Weyl-rescaled metric to make the 3D metric dimensionless in the Einstein frame as $g^{(3)}_{ij} \to (r^2/R^2)g_{ij}^{(3)}$.
For $R\to \infty$, the cosmological constant term dominates and it gives positive potential. For $R \lesssim 40\,{\rm eV}^{-1}$, the photon and graviton contribution makes the potential negative.
For $R \ll m_\nu^{-1}$, active neutrinos have more degrees of freedom than the photon plus graviton, and hence the potential again becomes positive.
Therefore, there appears an AdS minimum around the neutrino mass scale if there were no
additional light fermions. The presence of additional light fermions can drastically change the situation.

\begin{figure}
    \begin{center}
        \includegraphics[scale=0.6]{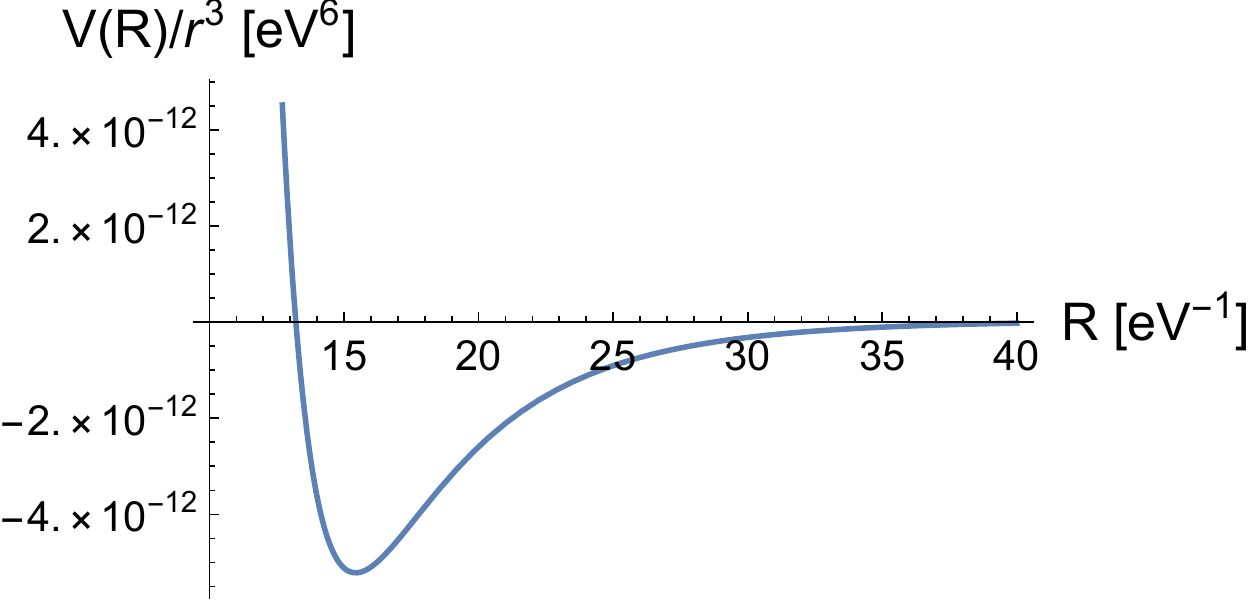}
        \includegraphics[scale=0.6]{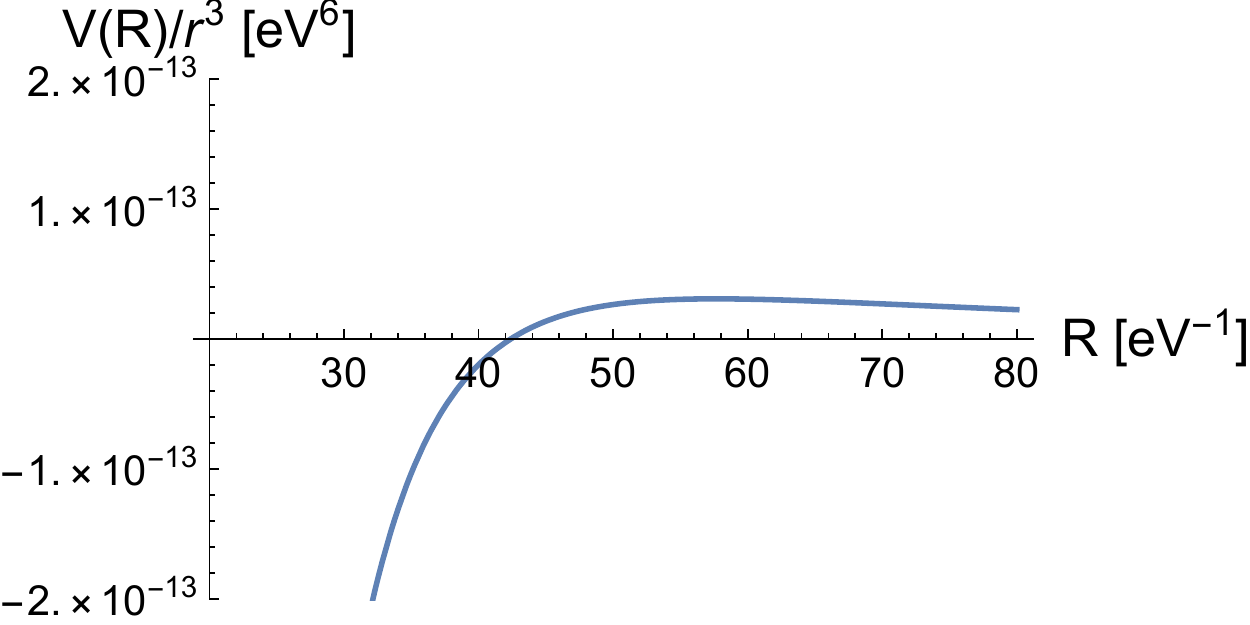}
        \caption{(Left) The radion potential for the compactified SM with three Majorana neutrinos where 
        the lightest one is taken to be massless.
        (Right) Magnification of the left panel.
        }
        \label{fig:potSM}
    \end{center}
\end{figure}

\begin{figure}
    \begin{center}
        \includegraphics[scale=0.6]{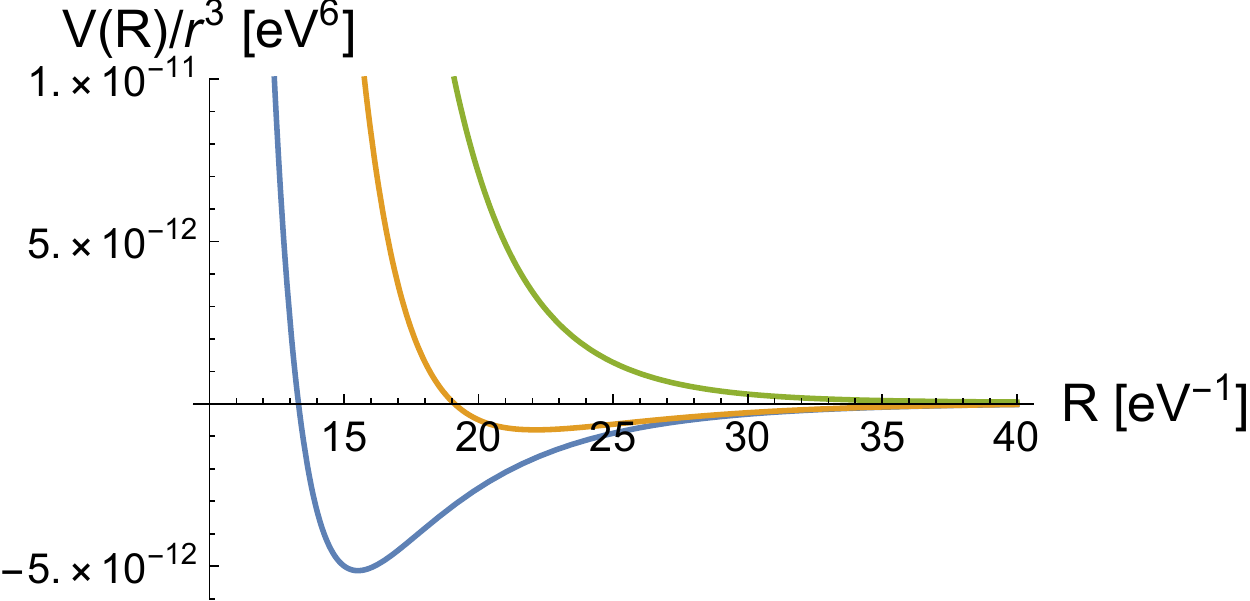}
        \includegraphics[scale=0.6]{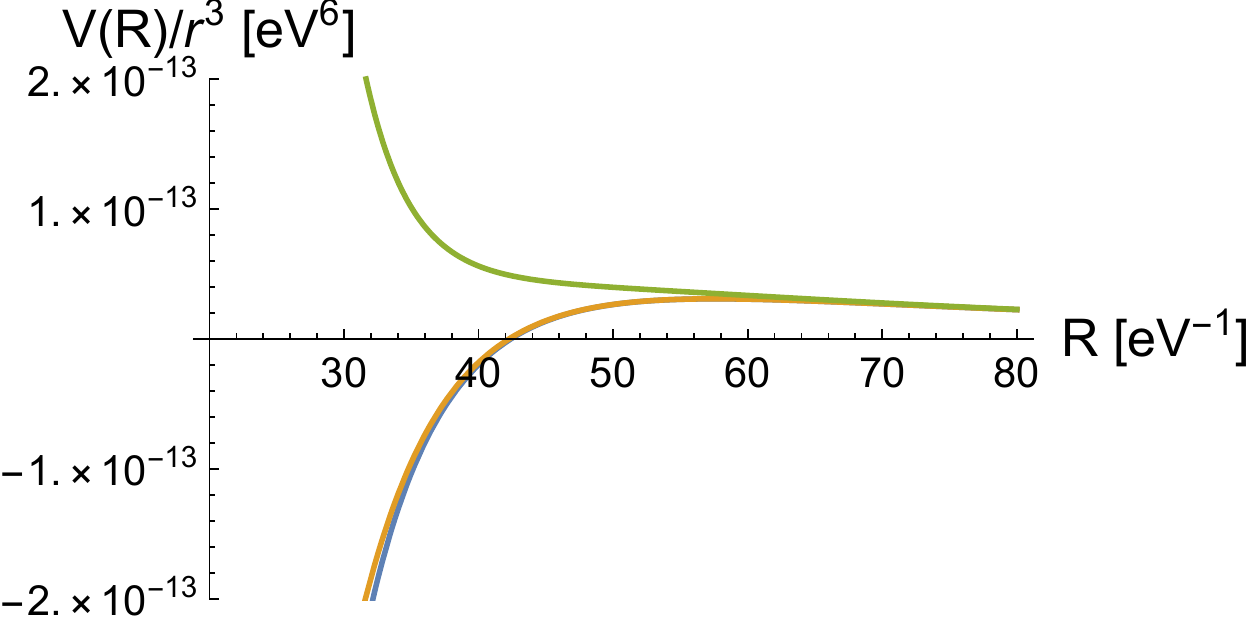}
        \caption{
        Same as Fig.~\ref{fig:potSM}, but adding        
        one extra Weyl fermion whose mass is $0.1\,{\rm eV}$, $0.03\,{\rm eV}$ and $0.01\,{\rm eV}$ from bottom to top.
        }
        \label{fig:pot}
    \end{center}
\end{figure}

Fig.~\ref{fig:potSM} shows the radion potential for the compactified SM with three Majorana neutrinos where
 the lightest one is taken to be massless.\footnote{
    The results are unchanged if the mass of the lightest active neutrino  is lighter than $\sim 0.01$\,eV.
} It is seen that there is an AdS minimum around the radius corresponding to the neutrino mass scale.
In Fig.~\ref{fig:pot}, we show the radion potential when we add a chiral fermion with mass of $0.1\,{\rm eV}$, $0.03\,{\rm eV}$ and $0.01\,{\rm eV}$ from bottom to top.
The AdS minimum disappears if the mass of the additional chiral fermion is smaller than $\sim 0.01$\,eV, which is consistent with the analysis in Ref.~\cite{Ibanez:2017kvh}.
Since our NTDM model predicts the existence of chiral fermion which is much lighter than $0.01$\,eV for $v_\Phi \lesssim \mathcal O(10^{14})$\,GeV
(see Eq.~(\ref{m5})), it ensures that there appears no 3D AdS minimum around the neutrino mass scale 
when the SM is compactified on a circle. This may be an appealing feature of our NTDM model 
in light of the sharpened WGC proposed in Ref.~\cite{Ooguri:2016pdq}.

Several remarks are in order.
In Ref.~\cite{Hamada:2017yji} it was pointed out that the radion potential in the SM  exhibits a runaway behavior toward the small radius if the minimization with respect to the Wilson line is taken into account. Still, there is a local minimum around the neutrino mass scale, and it is an open question whether there is a quantum tunneling from the local AdS minimum to the runaway vacuum. In this respect, Ref.~\cite{Buratti:2018onj} proposed that the WGC prohibit even the local AdS minima.  
Thus the erasure of the local AdS minimum around the neutrino mass scale, as in our NTDM model, may be important.
In Ref.~\cite{Gonzalo:2018tpb} more complicated geometry of the compactified dimension was considered, and it was found that there may be an AdS minimum around the QCD scale. Our NTDM scenario does not affect the vacuum structure around the QCD scale. However, it remains to be seen to what extent the WGC of Ref.~\cite{Ooguri:2016pdq} can be applied to various choices of the compactified geometry. 

\section{Conclusions}
In this Letter, we have revisited the NTDM scenario where one of the extra fermions that are introduced
to satisfy the anomaly cancellation conditions of \ubl explains the observed DM. In the case of 
an integer ${\rm B\,\mathchar`- L}$ charge assignment, there appear two light fermions in the low energy.
The lighter one is essentially decoupled from the others except for the \ubl gauge interactions, 
and it is stable due to the residual $Z_2^{\rm B\,\mathchar`- L}$ symmetry. On the other hand,
the heavier one has nonzero mixings with the other fermions and decays into  $\nu\nu\bar\nu$ and other modes depending
on its mass. We have explored a new parameter region where the heavier one becomes sufficiently long-lived
to be DM for $m_{\psi_\ell}  \lesssim {\cal O}(10)$\,keV or ${\cal O}(10)$\,MeV depending on the ${\rm B\,\mathchar`- L}$ charge assignment.
For a certain choice of the parameters, it can even explain
the tantalizing hint for the X-ray line excess at $3.55$\,keV. While the lighter one has no cosmological impact in this case, it can erase the AdS vacuum around the neutrino mass scale that appears in a compactification of the SM on a circle. 
Thus, the SM can be consistent with the weak gravity conjecture
that suggests the absence of AdS vacua, in our NTDM model.

\section*{Acknowledgments}

This work is supported by JSPS KAKENHI Grant Numbers JP15H05889 (F.T.), JP15K21733 (F.T.), 
JP26247042 (F.T),  JP17H02875 (F.T.), JP17H02878(F.T. and T.T.Y),
JP18K03609 (K.N.), JP15H05888 (K.N.), 17H06359 (K.N.), JP26104009 (K.N. and T.T.Y), JP16H02176 (T.T.Y),
and by World Premier International Research Center Initiative (WPI Initiative), MEXT, Japan.

\bibliography{reference}

\end{document}